\documentstyle[11pt,aaspp4]{article}
\def\puncspace{\ifmmode\,\else{\ifcat.\C{\if.\C\else\if,\C\else\if?\C\else%
\if:\C\else\if;\C\else\if-\C\else\if)\C\else\if/\C\else\if]\C\else\if'\C%
\else\space\fi\fi\fi\fi\fi\fi\fi\fi\fi\fi}%
\else\if\empty\C\else\if\space\C\else\space\fi\fi\fi}\fi}
\def\SP{\let\\=\empty\futurelet\C\puncspace }
\def\etal{et\SP al.\SP }
\def\kms{kms$^{-1}$}
\def\h-1{$h^{-1}$}

\def\h1{$h^{-1}$}
\def\etal{et al. \,}
\def\eg{e.g., \,}
\def\dlf {$\phi(M)$\SP}

\def\dunit{$\times 10^{-2}h^3$Mpc$^{-3}$}

\begin{document}

\title{Estimating Galaxy Luminosity  Functions}

\author{C.N.A. Willmer} 
\affil{Observat\'orio Nacional, Rua General Jos\'e Cristino
77, Rio de Janeiro, RJ 20921-030, Brazil} 
\affil{electronic mail: cnaw@on.br}

\begin{abstract}
In this work a comparison between different galaxy luminosity function
estimators by means of Monte Carlo simulations is presented. 
The simulations show that the  C$^-$ method of Lynden-Bell (1971) and
the STY method derived by Sandage, Tammann \& Yahil 
(1979) are the best estimators to measure the shape of the luminosity
function. The simulations also show that the STY estimator has a bias
such that the faint-end slope is underestimated for steeper
inclinations of the Schechter Function, and that this bias becomes
quite severe when the sample contains only a few hundred
objects. Overall, the C$^-$ is  
the most robust estimator, being less affected by different values of
the faint end slope of the Schechter parameterization and sample
size. The simulations are also used to compare different 
estimators of the luminosity function normalization. They demonstrate
that most methods bias the recovered mean density towards values which
are $\sim$ 20\% lower than the input value. 
\end{abstract}
 
\keywords{cosmology: observations -- galaxies: luminosity function
-- methods: numerical} 
\clearpage

\section {Introduction}
The luminosity function of galaxies is an important tool in the study of
large-scale structure, as it ultimately allows the estimation of the
total content of luminous matter in the form of galaxies (for a
comprehensive review on many applications of the
luminosity function the reader is referred to Binggeli, Sandage \&
Tammann 1988). As described by Binggeli et al. (1988), over the years
several methods have been proposed to calculate the luminosity
function and have been applied to a variety of samples of field (and
cluster) galaxies, as well as quasars.   
The first determinations of the luminosity function simply
counted the number of sample objects inside a given volume $\Phi =
N/V$, and were used in early works (\eg Hubble 1936), but, as noted by
Binggeli et al. (1988), detailed descriptions of the method only 
appeared much later (Trumpler \& Weaver 1953; Christensen 1975;
Schechter 1976). This has come to be known as the $classical$ $method$,
a term coined by Felten (1976), and is based on the assumption that the
distribution of galaxies in space is uniform (Binggeli \etal
1988). An estimator derived from the classical method was published by
Schmidt (1968) to study quasar evolution, which takes into account a
weight inversely proportional to the luminosity of the object. This is
known as the $1/V_{max}$ method, but as in the case of the classical
estimator, it contains the underlying assumption 
that the distribution is uniform. This estimator  was first 
applied to a sample of galaxies by Huchra \& Sargent (1973), while a 
paper describing how to combine different samples coherently was
presented by Avni \& Bahcall (1980). A further development using the
$1/V_{max}$ method was presented by Eales (1993) who used it to
calculate the luminosity function as a function of redshift.

In order to overcome the
dependence on the assumption of a uniform distribution, Lynden-Bell (1971)
derived the C$^-$ method, which was also applied to the sample of
quasars studied by Schmidt (1968). Lynden-Bell (1971) showed that the
C$^-$ estimator is a maximum-likelihood method, and requires no
assumption on the distribution, save that the luminosity
function is of the same shape at all points along the line of sight,
and that the sample should be ordered in 
luminosity. Further work on the C$^-$ was carried out by Jackson
(1974) who extended it so as to analyse several samples, where the
selection function of each sample is taken into account. By assuming
that the distribution of objects could be described by analytic
functions, Jackson (1974) also obtained error estimates for this method.
A similar method to the C$^-$ was proposed by Nicoll \& Segal (1983),
but differs in the sense that it uses binning, in contrast to
Lynden-Bell (1971), where no binning is used. The application of the
C$^-$ method to galaxies was first proposed by  Choloniewski (1987),
who by using simple arguments showed that it can give a properly
normalized luminosity function, without making any assumption on the
shape of the density or luminosity distributions. 

A method developed to remove effects caused by the density
inhomogeneities was derived by Turner (1979) and independently by
Kirshner, Oemler \& Schechter (1979), who considered the
ratio of the differential luminosity function at each absolute
magnitude interval between $M$ and $M + dM$ and the total number of
galaxies brighter than $M$. A slightly modified version of this method
binning galaxies in equal radial velocity intervals instead of equal
magnitude bins was used by Davis \& Huchra (1982, hereafter DH) and de
Lapparent, Geller \& Huchra (1989, hereafter LGH).

Another maximum-likelihood estimator was developed by Sandage, Tammann
\& Yahil (1979, hereafter STY), who applied it to the Revised Shapley
Ames catalog of galaxies (Sandage \& Tammann 1981). This estimator also
cancels out the contribution of the density distribution, and allows
including corrections for galaxy incompleteness or other effects. In
contrast to the  C$^-$, this method assumes that the luminosity
distribution is described by an analytic function.  In order to
provide a visual representation as well as an estimate of the
goodness-of-fit of the STY maximum-likelihood, Efstathiou, Ellis \&
Peterson (1988, hereafter EEP) derived the stepwise maximum likelihood
method (SWML)  where galaxies are counted in magnitude bins, but where no
functional shape is assumed. The stepwise maximum likelihood has been
extended by Heyl \etal (1997) to take into account the dependence of
the luminosity function 
with redshift, and by Springel \& White (1997) who instead of
assuming the non-parametric luminosity function as being described by a
constant value in each magnitude bin (as in a histogram), assume
power-laws so that the distribution becomes smoother.

Two further maximum-likelihood and clustering insensitive methods were
derived by Marshall \etal (1983) and Choloniewski (1986) by using as
a basic assumption the fact that the distribution of objects is the result of
a random process described by a Poisson probability distribution
(Binggeli \etal 1988). Marshall \etal (1983) assumed that the
luminosity and density distributions could be described by an analytic
form, and applied it to a sample of quasars. On the other hand,
Choloniewski's  (1986) derivation is non-parametric, but requires
binning the data, and was applied by its inventor to a sample of
galaxies taken from the CfA1 survey (Huchra \etal 1983). 

In spite of the variety of methods that have been proposed to
calculate the luminosity function (Binggeli \etal 1988 list 13 in
their table 2), no detailed comparison between them has been carried
out. For instance, although many possibilities have been forwarded to
explain the observed discrepancy between the normalizations
measured for the surveys of galaxies nearby and at intermediate
redshifts ($z \leq$ 0.1) (\eg da Costa \etal 1994; Marzke, Huchra \&
Geller 1994, hereafter MHG; Lin et al. 1996) relative to distant
samples (\eg Lilly \etal 1995; Ellis \etal 1996), it is not clear
whether this could not be partly due to the procedures used in the
calculation of the luminosity function. It is interesting to note that
the nearby surveys have relied on the combination of the STY
maximum-likelihood with the SWML, while the surveys of distant
galaxies have used primarily the 1/$V_{max}$ estimator.
The results from these works show that in general the deeper samples
present higher values of the normalization than local ones
and could imply in the existence either of density evolution or a
population of ``disappearing'' galaxies. Other explanations
such as a poor determination of the faint end of the local
luminosity function (Gronwall \& Koo 1995; Koo, Gronwall \& Bruzual
1993), have also been proposed and are supported by recent
determinations made by da Costa \etal (1997) who find a faint end
slope $\alpha \sim -1.2$ for the SSRS2, while an even steeper slope
$\alpha \sim -1.5$ has been determined by Sprayberry \etal (1997)
using a survey of low surface brightness galaxies.

The main motivation of this paper is to examine how well can some
of the different estimators measure the luminosity function in an apparent
magnitude-limited survey. 
This will be done by comparing the results
obtained applying these techniques to
simulated catalogs of galaxies, and will examine if the use of
different estimators could be a possible source of the
discrepancies obtained by the different redshift surveys. 
Comparisons between some of the methods have appeared previously
(Choloniewski 1986; EEP; LGH; Ratcliffe et al. 1997), while
discussions on the properties of some estimators have been presented
by Petrosian (1992) and Heyl et al. (1997). 

This paper is organized as follows: Section 2 will describe the estimators; in
Section 3 we describe the Monte Carlo simulations and the results
we obtain.  The application of the different estimators to a
sample of real galaxies is presented in Section 4, followed by the
conclusions in Section 5. 
\section {Estimators}
\subsection{STY estimator}

The first estimator we will describe is the parametric
maximum-likelihood estimator proposed by STY. This estimator is
insensitive to the presence of 
fluctuations in the sample (e.g., STY; MHG), and does not require binning the data, so that all the
information contained in the sample is preserved. In the STY
formalism, one defines the cumulative probability that a galaxy at
redshift $z$ will have a magnitude brighter than $M$ as
\begin{equation}
P(M,z) = {\int_{-\infty}^{M} \phi(M') \rho(z) f(m') dM' \over
\int_{-\infty}^{+\infty} \phi(M') \rho(z) f(m') dM' }
\end{equation}
where $\phi(M)$ is the differential luminosity function, $\rho(z)$
represents the redshift distribution at $z$ and $f(m')$ 
the catalog incompleteness, where $m' = M' + 5 $log$ \lbrack
D_L(z)\rbrack + 25 + K(z,T)$  
is the apparent magnitude; $D_L(z)$ is the luminosity
distance and $K(z,T)$ the cosmological $K$-correction which depends on
the redshift $z$ and morphological type $T$. In this notation, one may easily 
incorporate other effects such as sampling rates (Lin et al. 1996),
and the magnitude error (EEP; MHG). The probability 
density that a galaxy will be detected in a 
redshift survey is obtained by calculating the partial derivative of
$P(M,z)$ relative to $M$:
\begin{equation}
p(M_i,z_i) ={ \phi(M_i) \over \int_{-\infty}^{M_{faint(z_i)}} \phi(M) dM }.
\end{equation}
Thus this probability is directly proportional to the value of the
differential luminosity function at $M_i$ and inversely proportional
to the faintest absolute magnitude visible at $z_i$, $M_{faint}(z_i)$
(Heyl et al. 1997). 
One can then define the likelihood
which is the joint probability of all galaxies in the sample
belonging to the same parent distribution:
\begin{equation}
{\cal{ L}} = \prod_{i=1}^{N_g} p(M_i,z_i).
\end{equation}
To estimate the maximum likelihood, one must maximize this function
relative to the $p(M_i,z_i)$. To do this, one has to assume some
parameterization of the luminosity function, such as a Schechter
(1976) function, and maximizing the likelihood relative to the
parameters $M^*$ and $\alpha$. The errors can be estimated from the
error ellipsoid defined as
\begin{equation}
ln {\cal{ L}} = ln  {\cal{ L}}_{max} - { 1\over 2} \chi_\beta^2 (N)
\end{equation}
where $\chi_\beta^2 (N)$ is the $\beta$ point of the $\chi^2$
distribution with N degrees of freedom (EEP).
Because this method involves ratios between the differential and
integrated luminosity functions, and we are assuming that the
fluctuations are independent of $\phi (M)$, the normalization has to
be calculated independently.
\subsection{SWML} 
As discussed by EEP, the STY maximum likelihood presents the
inconvenience that one cannot test whether the 
parameterization represents a good fit to the data. In order to provide
an adequate visual representation of the data, EEP derived  the
Stepwise Maximum Likelihood method, which does not rely on the
assumption of a simple functional form for \dlf. This is done by
parameterizing \dlf as a series of  $N_p$ steps:

\begin{equation}
\phi(M) = \phi_k, \quad M_k-\Delta M/2 < M_k < M_k+\Delta M/2, \quad k= 1,
... , N_p.
\end{equation}

Let $x = M_i - M_k$. We can then define two window functions (EEP; MHG):
\begin{equation}
W(x)  = \left\{  \begin {array} {ll}
1, & \quad |x| \leq \Delta M/2 \\
0, & \quad |x|   >  \Delta M/2 \end{array} \right .
\end {equation}
so that $W(x)  \phi_k$ is the differential luminosity function at
$M_k$, and
\begin{equation}
H(x) = \left\{ \begin {array} {cc}
               1, & \quad x \leq -\Delta M/2 \\
{1 \over 2} - {x \over \Delta M}, & \quad -\Delta M/2 \leq x \leq +\Delta M/2 \\
               0, & \quad x \geq +\Delta M/2.\end{array} \right .
\end{equation}
One may then re-write the denonimator in Eq. (2) as: 
\begin{equation}
 \sum_{i=1}^{N_g}
\biggl\lbrace \sum_{j=1}^{N_p} \phi_j \Delta M H \lbrack  M_j -
M_{faint(z_i)}\rbrack \biggl\rbrace
\end{equation}
The likelihood function can then be written as
\begin{equation}
ln {\cal{ L}} = \sum_{i=1}^{N_g} W(M_i-M_k) ln\phi_k - \sum_{i=1}^{N_g}
\biggl\lbrace \sum_{j=1}^{N_p} \phi_j \Delta M H \lbrack  M_j -
M_{faint(z_i)}\rbrack \biggl\rbrace + constant
\end{equation}
The likelihood is calculated using an arbitrary normalization, and in
order to compare different surveys, EEP imposed an additional
constraint:
\begin{equation}
g =  \sum_{k=1}^{N_p} \phi_k \Delta M 10^{0.4 \beta (M_k-M_f)} - 1 = 0
\end{equation}
where $\beta$ is a constant, chosen by EEP to be $\sim$ -1.5 in order
to allow minimum variance, and $M_f$ is a fiducial magnitude. By
including this constant with a lagrangian multiplier, one defines
\begin{equation}
ln {\cal{ L}}' = ln {\cal{ L}} + \lambda g(\phi_k),
\end{equation}
which is maximized relative to the $\phi_k$ and $\lambda$ using
condition (10) and $\lambda$ = 0. The $\phi_k$ are obtained iteratively
from 
\begin{equation}
\phi_k \Delta M = \frac{ \displaystyle{\sum_{i=1}^{N_g} W(M_i-M_k) }}
{\displaystyle{ \sum_{i=1}^{N_g} \biggl\lbrace   H \lbrack M_k - M_{faint(z_i)} \rbrack /
  \sum_{j=1}^{N_p} \phi_j \Delta M H \lbrack M_j - M_{faint(z_i)}} \rbrack
\biggl\rbrace }
\end{equation}	
This method also allows incorporating effects such as the catalog
incompleteness (\eg Marzke \& da Costa 1997). Because of the
arbitrary normalization, the mean density must be estimated
independently. The errors for the $\phi_k$ may be estimated from the
covariance  matrix, which is the inverse of the information matrix
(EEP):
\begin{equation}
{\bf{cov}} (\phi_k) = \left\lbrack {\bf{I}}(\phi_k)\right\rbrack^{-1} = \left\lbrack \begin {array} {cc}
\frac {\partial^2 ln {\cal{ L}} } {\partial \phi_i \partial \phi_j } +
\frac {\partial g } {\partial \phi_i} \frac {\partial g } {\partial
\phi_j} & \frac {\partial g} {\partial \phi_j} \\
 \frac {\partial g} {\partial \phi_i} & 0 \end{array} \right\rbrack_{\phi =
\phi_k}^{-1}
\end{equation}

\subsection{Choloniewski Method} 
A similar method to the SWML was proposed by Choloniewski (1986). In
this method galaxies are binned in equal intervals of absolute magnitudes and
distance modulus ($\mu$). By assuming that the distribution of
galaxies in the $M-\mu$ plane is described by a Poisson process, the
probability of finding $N_{ij}$ galaxies in a cell of width $dM$ is
\begin{equation}
P(N_{ij}) = exp(-\lambda_{ij}) \frac {\lambda_{ij}^{N_{\scriptscriptstyle{ij}}}} {N_{ij} !}
\end{equation}
where 
\begin{equation}
\lambda_{ij} = {1 \over \bar n} \phi_i \Delta M \rho_j {\Omega \over 3}
( r_j^3 - r_{j-1}^3);
\end{equation}
$\bar n$ is the mean density of the sample, $\phi_i$ the mean value of the
luminosity function in the $\Delta M$ interval and $\rho_j$ the local
value of the galaxy density at a distance $r_j$, given by the distance
modulus $\mu_j$.
The likelihood for a magnitude-limited sample is then
\begin{eqnarray}
{\cal{ L}}   = & {\displaystyle{ \prod_{i=1}^A  \prod_{j=1}^B }} &  exp(- \lambda_{ij})
   \frac { \lambda_{ij}^{N_{ij}}} {N_{ij} }  \\
               & \scriptscriptstyle{ {i+j \leq S}} \nonumber   
\end{eqnarray}
where A and B are the number of bins in magnitudes and distance
modulus respectively and S represents the selection line given by
$S = M + \mu$, $\mu$ being the distance modulus of an object with
absolute magnitude $M$ when it is at the survey's 
limiting apparent magnitude $m_{lim}$. This expression  is formally
identical to Eq. (3) in the STY method, as we are also considering a
product of probabilities.  As 
described by Choloniewski (1986), this likelihood is maximized
relative to the parameters $E_k = (\bar n, \phi_i, \rho_j)$, and a
non-parametric estimate is obtained both for the luminosity function
and the density distribution. However, as pointed
out by Choloniewski (1986), this estimator does not use all
information available in the sample, even for objects which are
contained in the survey. The reason for this is that when one projects
the distribution of galaxies on the $M-\mu$ plane, only those cells
which are fully contained within the survey limits are used. Cells
which are at the survey limit and thus only partially contained are
ignored. One could in principle solve this by using bins as small as
possible, but, as discussed by Choloniewski (1986), smaller bins tend to give
more biased results and larger errors in the determination of the likelihood.
The maximum-likelihood is estimated iteratively and converges rather
rather fast. Once the maximum-likelihood
parameters are known, one may fit the non-parametric $\phi_k$
distribution with a parametric representation. The uncertainties for
the individual estimators may be calculated in a similar fashion as in
the SWML method, by means of the covariance matrix:
\begin{equation}
{\bf{cov}}(E_k) \approx \left\lbrack \frac { \partial^2 {\cal{ L}}} {\partial
E_i \partial E_j} \right\rbrack^{-1}_{E=E_k}
\end{equation}
This method differs from the STY and SWML in the sense that the mean
density, and therefore the normalization of the sample is estimated
simultaneously with the luminosity function.

\subsection{Turner Method} 
The Turner (1979) method uses the same assumptions as the STY
estimator, but differs in the way the luminosity function is
calculated. As in the STY method, for an absolute
magnitude $M'$, the expected number of objects in a survey is given by
\begin{equation}
\langle n(M') dM \rangle = \rho(M') \phi(M') dM
\end {equation}
where $\rho(M')$ is the density distribution.
The number of objects with absolute magnitudes brighter than $M'$ and
apparent magnitudes brighter than the survey limit is
\begin{equation}
\langle N(M') dM \rangle = \rho(M') \int_{-\infty}^{M'} \phi(M) dM.
\end{equation}
One then defines a quantity, where the density cancels out:
\begin{equation}
Y(M') dM = \frac{\langle n(M') dM \rangle}{\langle N(M') dM \rangle} =
\frac { \phi(M') dM} { \int_{-\infty}^{M'} \phi(M) dM} = \frac {d
\Psi} {\Psi},
\end {equation}
where $\Psi$ is the integral luminosity function. This expression is
identical to equation (2). One can make the approximation
\begin{equation}
Y(M') \approx \frac {d ln \Psi(M \leq M')} {dM}.
\end{equation}
The integration of this equation yields the integrated luminosity
function and the differential luminosity function can be derived
by re-writing equation (20) as $\phi(M) dM = \Psi(M) Y(M) dM$ (LGH) or:
\begin{equation}
\phi(M) \approx \phi_0 Y(M) exp \biggl\lbrace \int_{-\infty}^{M} Y(M) dM
 \bigg\rbrace
\end{equation}
where $\phi_0$ is an integration constant.

In practice, one
calculates $Y(M)$ in magnitude bins (LGH)
\begin{equation}
Y(M) \delta M = \frac {N (\leq M) - N( \leq M - \delta M)} { N (\leq
M)}
\end {equation}
which we represent as $\delta M$, as they may be linear in magnitudes
(Turner 1979; Kirshner, Oemler \& Schechter 1979), or a function
derived from linear bins in redshift (DH; LGH).
The errors associated to $Y(M)$ may be assumed as following a Poisson
distribution (\eg LGH).

\subsection{C$^-$ Method} 
The C$^-$ method was derived by Lynden-Bell (1971) and was extensively
analyzed by Jackson (1974), Choloniewski (1987) and SubbaRao \etal
(1996). The latter 
also adapted this method so it could be applied to a sample with
photometric redshifts.  The C$^-$ method is the limiting
case of the last three methods described above, when each bin contains one
only object. With the C$^-$ method one recovers the integral luminosity
function $\Psi (M_i)$, and the differential
luminosity function may be obtained through $\phi (M) = - d \Psi /
dM$. Following Lynden-Bell (1971) we call the observed sample at $M$
of $X(M)$. In 
general for a change $dM$ in the magnitudes, the change in the number
of observed objects is not the same as the change in the entire
distribution (Subbarao et al. 1996), because galaxies fainter than the
apparent magnitude limit are undetected:
\begin{equation}
\frac {d \Psi} { \Psi} > \frac {dX} {X}.
\end{equation}
However, one may construct $C(M)$, a subset of $X(M)$ where one 
makes the assumption that for an infinitesimal increase $dM$ the
variation in the observed distribution should be equal to the  change
in the overall distribution: 
\begin{equation}
\frac {dX} {C} = \frac {d \Psi} {\Psi}.
\end{equation}
The function $C(M)$ represents the total number of observed objects
with absolute magnitudes brighter than $M$.  Expression (25) can be
integrated to yield the integrated luminosity function
\begin{equation}
\Psi (M) = A exp \biggl\lbrace \int_{-\infty}^{M} \frac {dX} {C}
\biggl\rbrace = A exp \biggl\lbrace \int_{-\infty}^{M} \frac {1}
{C(M)} \frac {dX(M)} {dM} dM \biggl\rbrace ,
\end{equation}
which is similar to the expression obtained in the Turner method.

In practice, for a redshift survey one may write $\frac {dX(M)} {dM}$
as a series of Dirac delta functions. However, because at the points $M=M_i$,
$C(M)$ is indeterminate (Lynden-Bell 1971; Subbarao et al. 1996),
one must consider the following approximation (Lynden-Bell 1971) for
each $M_i$:
\begin{equation}
C(M) = C^{-}(M_i) + x
\end{equation}
where $x = X(M^{-}) - X(M)$ is the variation of the number of galaxies
in the interval $M^{-} - M$ and $C^{-}(M_i)$ is the total number of
objects with magnitudes brighter than $M_i$, but with the point $i$
itself omitted (Lynden-Bell 1971). One may then integrate
over the interval just around $M_i$:
\begin{equation}
\int_{M_{i\scriptscriptstyle{-}}}^{M_{i\scriptscriptstyle{+}}} \frac
{1} {C(M_i)} dX(M_i) = \int_0^1 \frac {dx} {C^{-} + x} = ln \left\{
\frac {C^{-}(M_i) + 1} {C^{-}(M_i)} \right\}.
\end{equation}
One may then rewrite (26) as
\begin{equation}
\Psi(M_k) = A exp  \sum_{j=1}^k ln \left\{ \frac {C^{-}(M_j) +
1} {C^{-}(M_j)} \right\} = A \prod_{j=1}^k \frac {C^{-}(M_j) +
1}{C^{-}(M_j)}.
\end{equation}
With this expression one may construct the non-parametric luminosity
function, while the integration constant $A$ will represent the
normalization of the luminosity function. The normalization constant
can be obtained from any non-zero number, and working up and down the
magnitude range using factors of $\lbrack C^{-}(M_i) + 1 \rbrack /
C^{-}(M_i)$ and its inverse (\eg Lynden-Bell 1971). Jackson (1974)
shows how the normalization can be obtained when several samples are
combined. This is done by determining the expected
number of objects in each subsample using an expression analogous to
(39) below, where each subsample is weighted by the appropriate
selection function. The normalization is then obtained
comparing the expected number with total number of observed
objects. Jackson (1974) also showed how to estimate errors for the
C$^-$ method, by assuming that the distribution may be described
analytically and then calculating the covariance matrix, in a completely
analogous way as the SWML and Choloniewski methods.

\subsection{1/V$_{max}$} 
The last method that will be described is the $1/V_{max}$ method which was
first published by Schmidt (1968). A detailed study of
this method was carried out by Felten (1976), who showed it is a
minimum-variance maximum-likelihood estimator. The method assumes that
for a given absolute magnitude
\begin{equation}
\Psi(M) = \sum_{j=1}^{N_g} \frac {1} {V_{max}(j)}
\end{equation}
where $V_{max}(j)$ is the volume corresponding to the maximum
distance galaxy $j$ could be observed, and still be included in the
sample. One may also calculate the differential luminosity function
using a variant of the ``classical'' estimator, by counting galaxies
in magnitude bins as done by DH:
\begin{equation}
\phi(M) dM = \frac {\displaystyle {\sum_{i=1}^{N_g} N(M - dM/2 \leq
M_i \leq M + dM/2)}}  {V_{max}(M)}.
\end{equation}
Felten (1976) demonstrated that even though this estimator is biased,
because one loses information on where about in the magnitude bin a
galaxy is located, 
for small enough intervals $dM$, it does provide a reasonable estimate
of the luminosity function. The $1/V_{max}$ estimator has one
advantage over some of methods described above, in the sense that it gives
simultaneously the shape and the normalization of the luminosity
function. However, this is also its major drawback, because it is very
sensitive to density fluctuations.

\section {Normalization of the Luminosity Function}

Most of the methods described in the previous section recover the
shape of the luminosity function, but, with the exception of the
$1/V_{max}$ and the Choloniewski estimators (and in special cases the
C$^{-}$ ) they do not provide any information about the normalization
($\phi^*$), which has to be estimated in an independent manner. This
quantity is related to the mean density  ($\bar n$) of the sample through:
\begin{equation}
\phi^* = \frac { \bar n} { \int_{M_{\scriptscriptstyle
{bright}}}^{M_{\scriptscriptstyle{faint}}} \phi(M) dM }
\end{equation}
where $M_{bright}$ and $M_{faint}$ are the brightest and faintest absolute
magnitudes considered in the survey. The mean density is obtained
correcting the redshift distribution by the selection function, which
describes the probability of a galaxy at redshift $z$ being included
in a survey (DH; EEP; LGH):
\begin{equation}
s(z) = \frac {\int_{M_{\scriptscriptstyle {bright}}
}^{min(M_{\scriptscriptstyle {max(zi)}},M_{\scriptscriptstyle
{faint}})} \phi(M) 
dM} {\int_{M_{bright}}^{M_{faint}} \phi(M) dM}.
\end{equation}
where $M_{max(z_i)}$ is the faintest absolute magnitude detected at
redshift $z_i$.
Various methods have been proposed to calculate the mean
density. DH derived a minimum-variance estimator:
\begin{equation}
n = \frac {\displaystyle{ \sum_{i=1}^{N_g} N_i(z_i) w(z_i)}}
{\int_{0}^{z_{max}} s(z) w(z) \frac {dV} {dz} dz }
\end{equation}
where $N_i(z_i)$ is the number of galaxies at redshift $z_i$ and the
weights are inversely proportional to the selection function and the
second moment of the two-point correlation function, which represents
the mean number of galaxies in excess of random around each galaxy out
to a distance $r$:
\begin{equation}
w(z_i) = \frac {1} { 1 + {\bar n} J_3 s(z)}, \quad J_3 = \int_0^r r^2
\xi(r) dr.
\end{equation}
In this case, the density is obtained through iteration, and can be
calculated either by counting galaxies in redshift bins (\eg MHG) or by
considering bins containing only 1 galaxy. If one sets
$w(z_i)=1$, expression (34) reduces to the $n_3$ estimator of DH:
\begin{equation}
n_3 = \frac {N_T} { \int_0^{z_{max}} s(z) dz}
\end{equation}
This estimator is fairly robust, but is also affected by
large-scale inhomogeneities in the foreground (DH). Another estimator
proposed by DH is
\begin{equation}
n_1 = \frac { \int_0^{zmax} { \frac {N(z)} {s(z)} dz}}
{ \int_0^{zmax} \frac {dV} {dz} dz}.
\end{equation}
$N(z)$ in this equation represents the number of galaxies luminous
enough to be included in a shell of width $dz$ at redshift $z$. This
procedure gives higher weights to the more distant points where the
selection function becomes less certain. DH solved this potential bias
by limiting the determination to the range where $s(z) \geq$ 0.1, while
an alternative to truncation at a given value of $s(z)$ was made by
LGH who calculated the derived $\phi^*$ from the median and
mean values of the $n_1$ and $n_3$ estimators calculated over the
entire depth of the survey. 

A slightly different estimator from the above was proposed by EEP,
where the contribution of the two-point correlation function is not taken into
account, and is the $n_1$ estimator at the limiting case of only one
galaxy per bin:
\begin{equation}
n_{eep} = \frac {1} {V} \sum_{i=1}^{N_g} \frac {1} {s(z_i)}.
\end{equation}

Finally, another estimator proposed by EEP is the normalization by
means of the observed number counts. To obtain the normalization,
one calculates the expected number of galaxies by integrating
the luminosity function over the whole surveyed volume for each
apparent magnitude bin: 
\begin{equation}
A(M) = \int_0^\infty dz \frac {dV} {dz} \int_{-\infty}^{M_{max(z)}}
\phi(M) dM = \phi^* I(m)
\end{equation}
and the value of $\phi^*$ is obtained by minimizing 
\begin{equation}
\sum_n \frac {\lbrack dN(m_n) - \phi^* dI(m_n)\rbrack^2} { \phi^*
dI(m_n)}
\end{equation}
where $dN(m_n)$ is the observed number of galaxies to the limiting
apparent magnitude $m_n$. Following EEP, the normalization is then:
\begin{equation}
\phi^*_c = \left\lbrack \frac {\sum_n \lbrack dN(m_n) \rbrack^2 /
dI(m_n) } {\sum_n dI(m_n) }\right\rbrack^{1/2}
\end{equation}

The uncertainties in the mean density have only been derived in the
case of the minimum-variance estimator (DH):
\begin{equation}
\frac {\delta {\bar n}} {\bar n} = \left\lbrack \frac {1}
{\int_{0}^{z_{max}} s(z) w(z) \frac {dV} {dz} dz } \right\rbrack^{1/2} 
\sim \left\lbrack \frac {J_3} {V} \right\rbrack^{1/2}
\end{equation}
and when calculating the uncertainty of $\phi_{min var}$  one should
also take into account the uncertainties in $\alpha$ and
$M^*$ (Lin et al. 1996). Alternatively, to estimate the error in the
density  one could use the dispersion around the mean or median
values of the density counted in shells. By using this procedure, LGH
estimated that the uncertainty in the determination of $\phi^*$ is $\sim$
25 \% for the CfA2 slices. 
\section{The Monte Carlo Simulations}

For the Monte Carlo simulations the transformation method (Press
et al. 1986) was used to generate random variates first in redshift and then
in magnitude. A homogeneous  redshift distribution is assumed, and
is obtained by calculating for a given interval in redshift
\begin{equation}
N(z) = \int_{z_1}^{z_2} \int_{-\infty}^{M_{max}(z')} \phi(M) {dV
\over dz'} dM dz',
\end{equation}
where $dV/dz'$ is the
volume element corrected for relativistic effects.  In this work we
use the $\phi(M)$ as parameterized by  Schechter (1976), which
expressed in magnitudes is
\begin{equation}
\phi(M) dM = 0.4\; ln 10\; \phi^* 10^{0.4(M^*-M)(\alpha+1)} exp\lbrace -
10^{0.4(M^*-M)(\alpha+1)} \rbrace dM
\end{equation}
where $\phi^*$ is the normalization, $M^*$ a
fiducial magnitude that characterizes the point where the growth of the
function changes from a power-law to an exponential slope, the
inclination of which is described by $\alpha$. Although other
parameterizations for the luminosity function have been proposed (\eg
Yahil et al. 1991; see also Binggeli et al. 1988) they will not be
considered here. 

The luminosity function was calculated for the interval --21.5 + 5 $log \;
h$ to --14.0 + 5 $log \; h$, where the Hubble parameter is defined as $h =
H_0/100$ $km s^{-1} Mpc^{-1}$, though when generating the samples, we
considered a bright cut at --22.5. A total of 1000 simulations were run using
as input parameters $\alpha_{in}$ = --0.7, --1.1 and --1.5 in order to
verify the 
sensitivity of the estimators to the inclination of the faint
slope. These values were chosen so as to encompass the range of slopes
measured by Marzke \& da Costa (1997) when fitting luminosity
functions for galaxies in different color ranges.  The values of $M^*$ =
--19.1 and $\phi^*$ = 0.02 were used throughout. The simulations were 
calculated for a survey with a similar geometry to the CfA1, which has
a solid angle of 2.66 steradians and an apparent magnitude limit of
14.5, and the number of galaxies varied according to the faint end
slope, going from about 1522 to 1734.

The recovered distribution from the Monte Carlo simulations of
$\alpha$ and $M^*$ for the case 
of $\alpha_{in}$=--1.1 are presented as histograms in Figures 1 and 2
respectively, while 
the median values and an estimate of the dispersion obtained in the
Monte Carlo simulations are presented in Table 1.
In general all methods recover quite well the input
values, with the exception of the $1/V_{max}$ estimators where the
recovered values are usually biased either in $\alpha$ or in $M^*$
depending on whether galaxies are binned in redshift or magnitudes.
It is immediately apparent that the STY and C$^-$
methods provide the best results. This comes as no surprise, given
that these methods use all information available in the sample, since no
kind of binning is required. Both recover the input $M^*$ but
bias $\alpha$ in different directions. That the STY is somewhat biased
had been noted by EEP, but this is within the dispersion value. The
C$^-$ presents a slightly larger dispersion than the STY, as can be seen in the
figures.

In Table 1 we also list median values for the distribution of $\alpha$
and $M^*$ for two other values of the faint end slope. The
corresponding histogram for $\alpha$ is shown in Fig. 3. In these
simulations $M^*$ was kept constant. 
For steeper slopes, most methods tend to be biased in different
degrees. The best result is obtained with the  C$^-$, as it recovers the
values closest to the input parameters. The STY presents a larger bias
than before, and is more than 1 $\sigma$ away from the input
value. Because of the correlation between $\alpha$ and $M^*$, the
latter is biased towards  fainter (larger) values. This result is
somewhat unexpected as a more inclined slope would mean more faint
galaxies and therefore a better constraint on the $\alpha$ value
(SubbaRao \etal 1996). For the shallower slope ($\alpha$ = --0.7), the
results are consistent with those found for $\alpha$ = --1.1.

The behavior of the SWML is also worth noting, as it is usually
coupled to the STY, in order to provide a visual representation of the
latter, and used to estimate the goodness of fit.
Here we apply a least-squares fit to the SWML distribution, as a means
of estimating the Schechter parameters, and whether the fit parameters
present a similar behavior to the STY. An 
inspection of Table 1 shows that it gives slightly different results, and
in Fig. 1 we can see that $\alpha$ recovered from these fits presents
a bi-modal distribution. This second peak occurs in samples for
which the faint end of the luminosity function is under-represented;
the SWML is the only estimator where such a behavior
is seen. As may be seen in Fig. 3 this secondary peak
gets less defined as one goes to shallower inclinations. All other
estimators, in spite of their biases, present a fairly 
symmetric distribution. A direct comparison between the fit parameters
of the SWML with the STY $\alpha$ and $M^*$ is shown in the panels of
Fig. 4. In both figures we plot as dashed lines the value of the input
parameters, while the solid line corresponds to equal values on each
axis. From the figure one can see that in most cases the SWML fits
present more inclined slopes and brighter $M^*$ than the STY maximum
likelihood, save those making up the secondary peak seen in Fig. 1.
The $M^*$ distribution presented in Fig. 4(b), and again there is a
systematic difference between both measurements.

As described above the C$^-$ method may be considered as a limiting
case of the SWML, where 
only one object per bin is considered. This can be seen by comparing
equation (12) (EEP's 2.12 in their paper) with equation (25) in
Choloniewski (1987), where the integral luminosity function (obtained
via the C$^-$ method) is averaged in small magnitude
intervals. In practice, both methods present different
results, which could be partly explained from the fact that the
procedure followed in the computation is different. The SWML is
obtained through an iterative procedure where one determines the
differential luminosity function, while the fits based on the C$^-$
method are made using the integrated luminosity function, which, because
it is cumulative, is also less sensitive to fluctuations. This is
particularly important at the faint end where the small number of
objects can affect the distribution, as in the case of the SWML.

In Table 2 we present values for the luminosity function normalization
$\phi^*$, obtained in the simulations using as input parameters $\alpha$ =
--1.1 and $M^*$=--19.1. In all cases, excepting for $\phi_c^*$
which is obtained  using equation (41), the values were
obtained by applying equation (32) to the various mean density
estimators.  Histograms of the distribution of $\phi^*$
are presented in Figure 5, where we show 
for the $1/V_{max}$  and Choloniewski methods the
normalization obtained by means of a 3-parameter
least-squares fits. For the STY, SWML, Turner and C$^-$ methods we
present the normalization using the minimum variance weighting
averaged over redshift bins. 
The use of this weighting on
Poisson samples is not entirely correct as the weighting function for
a random sample should be
be $w(r)=0$ as $J_3=0$. However, the values we
recover are not much different from what one measures for the other
density estimators. These are presented in Figure 6, where we show
the recovered densities but now only for the STY solutions, mainly
because this is the most commonly used method. 

Both Fig. 5 and Fig. 6 show that the $\phi^*$ estimators present
rather large fluctuations, yet all of these show a trend of
underestimating the input value. The minimum-variance estimators
present rather similar dispersions, and 
the estimator which seems to present the best results is that
derived from the number counts. It should be noted, that in spite of the
observed discrepancies, all estimators agree within $\sim$ 30\% of the
input value. 

So as to verify the effect that making different absolute magnitude
cuts could have on the mean density, the luminosity function was
calculated for 500 simulations using the cuts in absolute magnitude
-21.5 $\leq M \leq$ -14 and -20.0 $\leq M \leq$ -14. The latter limits
are the ones used by Ellis et al. (1996) in their study of the
evolution of the luminosity function, while the former are close to
the limits used by survey efforts such as the SSRS2 (da Costa et
al. 1997; 1994), the Durham/UKST galaxy Redshift Survey (Ratcliffe et
al. 1997) and Stromlo-APM (Loveday et al. 1992). In general, the
$\phi^*$ estimated for the smaller magnitude interval is higher, but
the difference for most estimators is of the order of 5\%,
demonstrating that the difference in the normalization measured
between these works cannot be attributed to the different absolute
magnitude ranges.

In order to estimate the sensitivity of the estimators to the number
of galaxies in the sample, a series of simulations emulating the
survey of Willmer \etal (1994; 1996) were calculated. The latter aimed at
measuring redshifts of galaxies down to $b_J$ = 20 in a $4^\circ \times
0.67^\circ$ slice close to the North Galactic pole. In the simulations
the same Schechter parameters as above were used, and no
evolution was considered. The results from these simulations are
presented in Table 3, while the distribution of recovered $\alpha$
values is shown in Fig. 7. For this smaller sample of galaxies we find
that not only do the uncertainties become larger, but for some
methods, in particular STY, the biases increase, although this
increase also depends on the faint-end slope. For shallower slopes,
STY recovers the input parameters very well, while for steeper slopes,
it is more than 6 $\sigma$ off, and $M^*$ 2 $\sigma$ away from the
input values. The C$^-$ is also biased, and this bias tends to get
smaller as the slope increases, a behavior opposite to the STY. The
bias for $M^*$ is rather constant for the values of $\alpha$ we
considered. The SWML is less affected than STY at steeper slopes and
overall has a similar behavior to the  C$^-$ method, though the
magnitude of its
bias is always larger. The remaining estimators generally produce
better fits when the slopes are steeper, yet more than 1
$\sigma$ away from the input value. The Turner method in particular is
extremely sensitive to the number of objects as well as the bin width
used in the calculation. Finally, it is interesting to note
that for the smaller sample the Choloniewski and 1/$V_{max}$ (in
magnitudes) estimators give comparable results.

\section {Application of the estimators to the CfA1 Survey}

As an illustration of the behavior of the different estimators, and
to compare the results here with previous determinations, we have
calculated the luminosity function on a sample of real galaxies. For
this we chose the CfA1 survey, (Huchra et al. 1983; 1992), mainly because it
has been the sample most extensively studied with different
estimators.  In Table 4 we compare our measurements with those by
other authors, using the sample limits and bin widths (where
applicable) as stated in the original papers, which are indicated in
the references column. Here it should be noted that all determinations
excepting  Choloniewski (1986) used galaxies in both galactic
caps. The latter only considered galaxies in the northern galactic cap
portion of the CfA1 survey. Whenever possible, we compared the results
obtained by other authors for samples without corrections due to the
Virgo cluster. The results here show a reasonable agreement with most
of the previous determinations, usually within 0.1 both 
in $\alpha$ and $M^*$, save in the case of the STY (EEP) method as
calculated by EEP and the Turner method using redshift bins of
400 kms$^{-1}$ (DH). The origin of the discrepancy with EEP is unclear,
but could be due to the fact that they have taken into account the
Malmquist bias introduced by the errors of
Zwicky magnitudes. Another source of discrepancy is certainly the
catalog itself, as the version that was used here (nz40.dat dating from May
1994, see Huchra 1997) is more recent that any of the papers we
compare with. 

Another comparison between the different estimators is shown in Tables
5 (for $M^*$ and $\alpha$) and 6 (for $\phi^*$), again using the CfA1
survey, but now considering for all estimators the absolute magnitude range
-21.5 + 5 $log$ $h$ to -14.0 +  5 $log$ $h$, and as above, with no
corrections for Virgo, but correcting for the Local Group motion
following Yahil, Tammann \& Sandage (1977).
The sample contains 2395 galaxies, though in the
absolute magnitude interval we considered there are only 2345; all
galaxies with zero or negative heliocentric velocities or with corrected
velocities less than zero were removed.
A plot showing the different fits is presented in Fig. 8, and the
inset shows the estimated 1 $\sigma$ error ellipses. In the case of
the STY, Turner and C$^-$ methods, the differential luminosity
function is not calculated, so no observed distribution is
presented. From Table 5 one can see that save for the 
1/$V_{max}$, all estimates give fairly consistent results with
$\alpha$ = -1.2 and $M^*$= -19.2. 
It is interesting to note that now the STY values are fairly close to
the determination of EEP, though the comparison might not be fair, as
here there has been no correction for the magnitude errors. The values
for $\phi^*$ present a fairly large dispersion, with a larger
concentration of values close to 0.025 $h^3$ Mpc$^{-3}$. It is
interesting to note that the density measured using the 1/$V_{max}$
method are lower by a factor of 2 than the estimates using STY.

\section {Summary}

In this work several estimators for the luminosity function assuming a
Schechter form are compared. As would be expected, the methods
which use all information available in the sample (such as the STY and
C$^-$ methods) give the best results. It is also shown that even for
homogeneous samples the $1/V_{max}$ method using binning gives biased
results and in general tends to give higher values for the faint end slope than
any of the other estimators. The result of EEP that the STY
estimator is slightly biased is confirmed. The STY fit tends to
$underestimate$ the 
faint-end slope, giving flatter slopes for a fixed value of $M^*$.
This bias becomes worse for more inclined slopes and smaller samples.
A comparison between the least-squares fits to the SWML and the STY
maximum likelihood fits suggests that at more inclined values of the
faint-end slope, the goodness of fit estimated for STY could be
worsened because of the biases of both methods, which tend to grow in opposite
directions. This  result suggests that
for samples with steeper slopes another estimator, such as the C$^-$,
could be more adequate to calculate the non-parametric distribution
and fit.
We also find that the value for the mean density recovered
by most estimators are lower than the input values by factors ranging
up to 25 \%.  This implies that the observed discrepancy between
samples of local and distant galaxies cannot be attributed to the
different estimators used in the analyses carried out by the various
groups.

\bigskip\bigskip
\noindent 
I would like to thank D.C. Koo for initially proposing this project,
and L. da Costa, G. Galaz, D. Garcia-Lambas, C. Lobo, M. Maia,
R.O. Marzke and P.S. Pellegrini for discussions and suggestions, and
D. Lynden-Bell for his remarks on the history of the C$^-$
method. I also thank the anonymous referee whose comments have
helped to improve the presentation of this paper. This work has been partly
supported by CNPq grants 301364/86-9, 453488/96-0 and the ESO Visitor
Program.

\newpage
\makeatletter
\def\jnl@aj{AJ}
\ifx\revtex@jnl\jnl@aj\let\tablebreak=\nl\fi
\makeatother
\def\puncspace{\ifmmode\,\else{\ifcat.\C{\if.\C\else\if,\C\else\if?\C\else%
\if:\C\else\if;\C\else\if-\C\else\if)\C\else\if/\C\else\if]\C\else\if'\C%
\else\space\fi\fi\fi\fi\fi\fi\fi\fi\fi\fi}%
\else\if\empty\C\else\if\space\C\else\space\fi\fi\fi}\fi}
\def\SP{\let\\=\empty\futurelet\C\puncspace }
\def\etal{et\SP al.\SP }
\def\puncspace{\ifmmode\,\else{\ifcat.\C{\if.\C\else\if,\C\else\if?\C\else%
\if:\C\else\if;\C\else\if-\C\else\if)\C\else\if/\C\else\if]\C\else\if'\C%
\else\space\fi\fi\fi\fi\fi\fi\fi\fi\fi\fi}%
\else\if\empty\C\else\if\space\C\else\space\fi\fi\fi}\fi}
\def\SP{\let\\=\empty\futurelet\C\puncspace }
%
\tablenum{1}
\begin{deluxetable}{lcccccc}
\small
\tablewidth{0pc}
\tablecaption{Median values for recovered parameters $M^*$ and
$\alpha$, CfA1 like sample}
\tablehead{
\colhead{Method}    & 
\colhead{$\alpha$}  &
\colhead{M$^*$ }    &
\colhead{$\alpha$}  &
\colhead{M$^*$ }    &
\colhead{$\alpha$}  &
\colhead{M$^*$ }    
}
\startdata 
input values              & -1.10             &  -19.10 &-1.50  &  -19.10  &  -0.70  & -19.10 \nl
SWML                      & -1.13 $\pm$ 0.14  & -19.19 $\pm$ 0.10 & -1.45 $\pm$ 0.14  &  -19.13 $\pm$ 0.11  &   -0.82 $\pm$    0.14  &  -19.19 $\pm$    0.12 \nl
STY                       & -1.08 $\pm$ 0.07  & -19.10 $\pm$ 0.07 & -1.43 $\pm$ 0.06  &  -19.06 $\pm$ 0.07  &   -0.69 $\pm$    0.08  &  -19.10 $\pm$    0.06 \nl
Choloniewski              & -1.18 $\pm$ 0.11  & -19.04 $\pm$ 0.10 & -1.50 $\pm$ 0.08  &  -19.00 $\pm$ 0.11  &   -0.87 $\pm$    0.15  &  -19.08 $\pm$    0.12 \nl
Turner                    & -1.09 $\pm$ 0.11  & -19.13 $\pm$ 0.10 & -1.51 $\pm$ 0.09  &  -19.16 $\pm$ 0.10  &   -0.66 $\pm$    0.11  &  -19.11 $\pm$    0.09 \nl
$1/V_{max}$$_{\Delta z}$  & -1.31 $\pm$ 0.06  & -19.15 $\pm$ 0.08 & -1.67 $\pm$ 0.05  &  -19.14 $\pm$ 0.09  &   -0.94 $\pm$    0.08  &  -19.14 $\pm$    0.07 \nl
$1/V_{max}$$_{\Delta M}$  & -1.13 $\pm$ 0.07  & -18.99 $\pm$ 0.07 & -1.50 $\pm$ 0.06  &  -18.98 $\pm$ 0.07  &   -0.83 $\pm$    0.11  &  -19.04 $\pm$    0.09 \nl
C$^-$                     & -1.12 $\pm$ 0.09  & -19.10 $\pm$ 0.08 & -1.51 $\pm$ 0.07  &  -19.11 $\pm$ 0.08  &   -0.72 $\pm$    0.12  &  -19.10 $\pm$    0.07 \nl
\enddata
\end{deluxetable}

%
%
\makeatletter
\def\jnl@aj{AJ}
\ifx\revtex@jnl\jnl@aj\let\tablebreak=\nl\fi
\makeatother
\def\puncspace{\ifmmode\,\else{\ifcat.\C{\if.\C\else\if,\C\else\if?\C\else%
\if:\C\else\if;\C\else\if-\C\else\if)\C\else\if/\C\else\if]\C\else\if'\C%
\else\space\fi\fi\fi\fi\fi\fi\fi\fi\fi\fi}%
\else\if\empty\C\else\if\space\C\else\space\fi\fi\fi}\fi}
\def\SP{\let\\=\empty\futurelet\C\puncspace }
\def\etal{et\SP al.\SP }
\def\puncspace{\ifmmode\,\else{\ifcat.\C{\if.\C\else\if,\C\else\if?\C\else%
\if:\C\else\if;\C\else\if-\C\else\if)\C\else\if/\C\else\if]\C\else\if'\C%
\else\space\fi\fi\fi\fi\fi\fi\fi\fi\fi\fi}%
\else\if\empty\C\else\if\space\C\else\space\fi\fi\fi}\fi}
\def\SP{\let\\=\empty\futurelet\C\puncspace }
\def\dunit{$\times 10^{-2}h^3$Mpc$^{-3}$}
\tablenum{2}
\begin{deluxetable}{lccccccc}
\small
\tablewidth{0pc}
\tablecaption{Median values for the Schechter function normalization}
\tablehead{
\colhead{Estimator}    & 
\colhead{SWML}  &
\colhead{STY}   &   
\colhead{Choloniewski}  &
\colhead{Turner} &
\colhead{1/V$_{max}$($\Delta$ z) } &
\colhead{1/V$_{max}$($\Delta$ M) } &
\colhead{C$^-$}  \nl
 {} & {}   & {}  & {}   & \dunit  & {}   & {}  & {}
}
\startdata
$n$ ($\Delta z$)     & 1.41$\pm$0.33  & 1.68$\pm$0.21 & \nodata  & 1.47$\pm$0.26  & \nodata & \nodata & 1.85$\pm$0.25 \nl
$n$                  & 1.30$\pm$0.29  & 1.46$\pm$0.19 & \nodata  & 1.41$\pm$0.23  & \nodata & \nodata & 1.81$\pm$0.22 \nl
$n_1$   ($\Delta z$) & 1.69$\pm$0.28  & 1.85$\pm$0.21 & \nodata  & 1.77$\pm$0.26  & \nodata & \nodata & 2.10$\pm$0.26 \nl
$n_3$   ($\Delta z$) & 1.72$\pm$0.22  & 1.84$\pm$0.18 & \nodata  & 1.79$\pm$0.23  & \nodata & \nodata & 1.70$\pm$0.25 \nl
$n_3$                & 1.87$\pm$0.28  & 2.14$\pm$0.21 & \nodata  & 2.03$\pm$0.28  & \nodata & \nodata & 1.47$\pm$0.22 \nl
$n_{EEP}$            & 1.39$\pm$0.71  & 1.80$\pm$0.88 & \nodata  & 1.48$\pm$0.87  & \nodata & \nodata & 1.89$\pm$1.03 \nl
$\phi_c^*$           & 1.76$\pm$0.25  & 1.97$\pm$0.20 & \nodata  & 1.87$\pm$0.26  & \nodata & \nodata & 1.93$\pm$0.24 \nl
fit                  &  \nodata    & \nodata & 1.34$\pm$0.30&\nodata & 1.39$\pm$0.19 &  1.67$\pm$0.19 & 1.80$\pm$0.95 \nl
\enddata
\end{deluxetable}
%
%
\makeatletter
\def\jnl@aj{AJ}
\ifx\revtex@jnl\jnl@aj\let\tablebreak=\nl\fi
\makeatother
\def\puncspace{\ifmmode\,\else{\ifcat.\C{\if.\C\else\if,\C\else\if?\C\else%
\if:\C\else\if;\C\else\if-\C\else\if)\C\else\if/\C\else\if]\C\else\if'\C%
\else\space\fi\fi\fi\fi\fi\fi\fi\fi\fi\fi}%
\else\if\empty\C\else\if\space\C\else\space\fi\fi\fi}\fi}
\def\SP{\let\\=\empty\futurelet\C\puncspace }
\def\etal{et\SP al.\SP }
\def\puncspace{\ifmmode\,\else{\ifcat.\C{\if.\C\else\if,\C\else\if?\C\else%
\if:\C\else\if;\C\else\if-\C\else\if)\C\else\if/\C\else\if]\C\else\if'\C%
\else\space\fi\fi\fi\fi\fi\fi\fi\fi\fi\fi}%
\else\if\empty\C\else\if\space\C\else\space\fi\fi\fi}\fi}
\def\SP{\let\\=\empty\futurelet\C\puncspace }
\tablenum{3}
\pagestyle{empty}
\tablecaption{Median values for the Schechter function parameters,
smaller sample}
\begin{deluxetable}{lcccccc}
\small
\tablewidth{0pc}
\tablehead{
\colhead{Method}    & 
\colhead{$\alpha$}  &
\colhead{M$^*$}     &
\colhead{$\alpha$}  &
\colhead{M$^*$}     &
\colhead{$\alpha$}  &
\colhead{M$^*$ }    
}
\startdata 
input values              &         -1.10        &          -19.10     &
     -1.50    & -19.10  &  -0.70   & -19.10  \nl
SWML                      &   -1.18 $\pm$ 0.13  &  -19.24 $\pm$ 0.17 & -1.46 $\pm$ 0.10 & -19.12 $\pm$ 0.19 &   -0.94 $\pm$    0.17 &  -19.30 $\pm$    0.17 \nl
STY                       &   -1.06 $\pm$ 0.10  &  -19.07 $\pm$ 0.11 & -1.33 $\pm$ 0.07 & -18.94 $\pm$ 0.12 &   -0.69 $\pm$    0.12 &  -19.11 $\pm$    0.11 \nl
Choloniewski              &   -1.24 $\pm$ 0.14  &  -19.12 $\pm$ 0.19 & -1.51 $\pm$ 0.11 & -19.02 $\pm$ 0.19 &   -0.98 $\pm$    0.19 &  -19.20 $\pm$    0.19 \nl
Turner                    &   -0.82 $\pm$ 0.23  &  -19.07 $\pm$ 0.31 & -1.25 $\pm$ 0.20 & -19.12 $\pm$ 0.30 &   -0.78 $\pm$    0.21 &  -19.20 $\pm$    0.25 \nl
$1/V_{max}$$_{\Delta z}$  &   -1.29 $\pm$ 0.07  &  -19.40 $\pm$ 0.15 & -1.65 $\pm$ 0.07 & -19.42 $\pm$ 0.17 &   -1.06 $\pm$    0.10 &  -19.49 $\pm$    0.15 \nl
$1/V_{max}$$_{\Delta M}$  &   -1.21 $\pm$ 0.08  &  -19.08 $\pm$ 0.12 & -1.58 $\pm$ 0.07 & -19.06 $\pm$ 0.14 &   -0.93 $\pm$    0.14 &  -19.13 $\pm$    0.13 \nl
C$^-$                     &   -1.16 $\pm$ 0.13  &  -19.15 $\pm$ 0.14 & -1.53 $\pm$ 0.09 & -19.14 $\pm$ 0.14 &   -0.79 $\pm$    0.16 &  -19.15 $\pm$    0.14 \nl
\enddata
\end{deluxetable}
%
%
\makeatletter
\def\jnl@aj{AJ}
\ifx\revtex@jnl\jnl@aj\let\tablebreak=\nl\fi
\makeatother
\def\puncspace{\ifmmode\,\else{\ifcat.\C{\if.\C\else\if,\C\else\if?\C\else%
\if:\C\else\if;\C\else\if-\C\else\if)\C\else\if/\C\else\if]\C\else\if'\C%
\else\space\fi\fi\fi\fi\fi\fi\fi\fi\fi\fi}%
\else\if\empty\C\else\if\space\C\else\space\fi\fi\fi}\fi}
\def\SP{\let\\=\empty\futurelet\C\puncspace }
\def\etal{et\SP al.\SP }
\def\puncspace{\ifmmode\,\else{\ifcat.\C{\if.\C\else\if,\C\else\if?\C\else%
\if:\C\else\if;\C\else\if-\C\else\if)\C\else\if/\C\else\if]\C\else\if'\C%
\else\space\fi\fi\fi\fi\fi\fi\fi\fi\fi\fi}%
\else\if\empty\C\else\if\space\C\else\space\fi\fi\fi}\fi}
\def\SP{\let\\=\empty\futurelet\C\puncspace }
\def\dunit{$\times 10^{-2}h^3$Mpc$^{-3}$}

\tablenum{4}
\begin{deluxetable}{lllllllll}
\small
\tablewidth{0pc}
\tablecaption{Schechter function parameters for CfA1 survey}
\tablehead{
\colhead{Method}              & \colhead{$M_{bright}$}    &
\colhead{$M_{faint}$}         &
\colhead{$\alpha$}            & \colhead{M$^*$ }          &
\colhead{reference}           & \colhead{$\alpha_{here}$} &
\colhead{M$^*_{here}$ }       & \colhead{Notes } 
}
\startdata 
STY         & -21.5 & -16.0  & -1.08 & -19.10& EEP & -0.83 &-19.00 & {} \nl
$1/V_{max}$ & -21.5 & -16.0  & -1.5  & -19.5 & DH  & -1.49 &-19.28 &$\Delta M$ = 0.2 \nl
Turner      & -21.5 & -16.0  & -0.9  & -19.2 & DH  & -1.04 &-19.27 &$\Delta v$ = 400 kms$^{-1}$ \nl
Turner      & -21.0 & ...    & -1.2  & -19.2 & LGH & -1.10 &-19.30 &$\Delta v$ = 700 kms$^{-1}$ \nl
Choloniewski& -21.5 & -16.0  & -1.09 & -19.2 & Choloniewski (1986) & -1.09 &-19.07 &$\Delta M$ = 0.25 \nl
\enddata
\end{deluxetable}
%
%
\makeatletter
\def\jnl@aj{AJ}
\ifx\revtex@jnl\jnl@aj\let\tablebreak=\nl\fi
\makeatother
\def\puncspace{\ifmmode\,\else{\ifcat.\C{\if.\C\else\if,\C\else\if?\C\else%
\if:\C\else\if;\C\else\if-\C\else\if)\C\else\if/\C\else\if]\C\else\if'\C%
\else\space\fi\fi\fi\fi\fi\fi\fi\fi\fi\fi}%
\else\if\empty\C\else\if\space\C\else\space\fi\fi\fi}\fi}
\def\SP{\let\\=\empty\futurelet\C\puncspace }
\def\etal{et\SP al.\SP }
\def\puncspace{\ifmmode\,\else{\ifcat.\C{\if.\C\else\if,\C\else\if?\C\else%
\if:\C\else\if;\C\else\if-\C\else\if)\C\else\if/\C\else\if]\C\else\if'\C%
\else\space\fi\fi\fi\fi\fi\fi\fi\fi\fi\fi}%
\else\if\empty\C\else\if\space\C\else\space\fi\fi\fi}\fi}
\def\SP{\let\\=\empty\futurelet\C\puncspace }
\def\kms{kms$^{-1}$}
\def\dunit{$\times 10^{-2}h^3$Mpc$^{-3}$}
\tablenum{5}
\begin{deluxetable}{llll}
\small
\tablewidth{0pc}
\tablecaption{Schechter function parameters for CfA1 survey, using
same limits for all methods}
\tablehead{
\colhead{Method}              & \colhead{$\alpha$}        &
\colhead{M$^*$ }              & \colhead{Notes } \nl
{} & {} & {} & { }
}

\startdata
SWML          & -1.20 $\pm$ 0.03  & -19.30 $\pm$ 0.04 & $\Delta M$ = 0.25\nl
STY           & -1.11 $\pm$ 0.08  & -19.17 $\pm$ 0.08 & { } \nl
Choloniewski  & -1.18 $\pm$ 0.05  & -19.26 $\pm$ 0.07 & $\Delta M$ = 0.25 \nl				       
Turner        & -1.11 $\pm$ 0.06  & -19.32 $\pm$ 0.05 & $\Delta z$ = 500 \kms \nl
1/V$_{max}$   & -1.59 $\pm$ 0.04  & -19.43 $\pm$ 0.07 & $\Delta M$ = 0.25\nl     			       
1/V$_{max}$   & -1.70 $\pm$ 0.05  & -19.55 $\pm$ 0.05 & $\Delta z$ = 500 \kms\nl				       
C$^-$         & -1.20 $\pm$ 0.01  & -19.21 $\pm$ 0.01 & { } \nl
\enddata
\end{deluxetable}
%
%
\makeatletter
\def\jnl@aj{AJ}
\ifx\revtex@jnl\jnl@aj\let\tablebreak=\nl\fi
\makeatother
\def\puncspace{\ifmmode\,\else{\ifcat.\C{\if.\C\else\if,\C\else\if?\C\else%
\if:\C\else\if;\C\else\if-\C\else\if)\C\else\if/\C\else\if]\C\else\if'\C%
\else\space\fi\fi\fi\fi\fi\fi\fi\fi\fi\fi}%
\else\if\empty\C\else\if\space\C\else\space\fi\fi\fi}\fi}
\def\SP{\let\\=\empty\futurelet\C\puncspace }
\def\etal{et\SP al.\SP }
\def\puncspace{\ifmmode\,\else{\ifcat.\C{\if.\C\else\if,\C\else\if?\C\else%
\if:\C\else\if;\C\else\if-\C\else\if)\C\else\if/\C\else\if]\C\else\if'\C%
\else\space\fi\fi\fi\fi\fi\fi\fi\fi\fi\fi}%
\else\if\empty\C\else\if\space\C\else\space\fi\fi\fi}\fi}
\def\SP{\let\\=\empty\futurelet\C\puncspace }
\def\dunit{$\times 10^{-2}h^3$Mpc$^{-3}$}
  
\tablenum{6}
\begin{deluxetable}{lccccccc}
\small
\tablewidth{0pc}
\tablecaption{Schechter function normalization estimates for CfA1}
\tablehead{
\colhead{Estimator}    & 
\colhead{SWML}  &
\colhead{STY}   &   
\colhead{Choloniewsky}  &
\colhead{Turner} &
\colhead{1/V$_{max}$($\Delta$ z) } &
\colhead{1/V$_{max}$($\Delta$ M) } &
\colhead{C$^-$} \nl
 {} & {}   & {}  & {}   & \dunit  & {}   & {}  & {}
}
\startdata
$n$ ($\Delta z$)     & 1.98   & 2.60   & \nodata & 1.74   &\nodata  & \nodata & 2.52  \nl
$n$                  & 1.78   & 2.32   & \nodata & 1.58   &\nodata  & \nodata & 2.25  \nl
$n_1$   ($\Delta z$) & 2.31   & 3.11   & \nodata & 2.06   &\nodata  & \nodata & 2.97  \nl
$n_3$   ($\Delta z$) & 2.18   & 2.63   & \nodata & 2.13   &\nodata  & \nodata & 2.46  \nl
$n_3$                & 2.52   & 3.04   & \nodata & 2.48   &\nodata  & \nodata & 2.82  \nl
$n_{EEP}$            & 1.81   & 2.88   & \nodata & 1.47   &\nodata  & \nodata & 2.72  \nl
$\phi_c^*$           & 2.76   & 3.32   & \nodata & 2.72   &\nodata  & \nodata & 3.09  \nl
fit                  &\nodata &\nodata & 1.36 	 &\nodata & 1.00    &1.16     & 2.37  \nl
\enddata                 
\end{deluxetable}        
\clearpage               

%
%
\newpage
\begin{figure}
\vspace{200mm}
\includegraphics{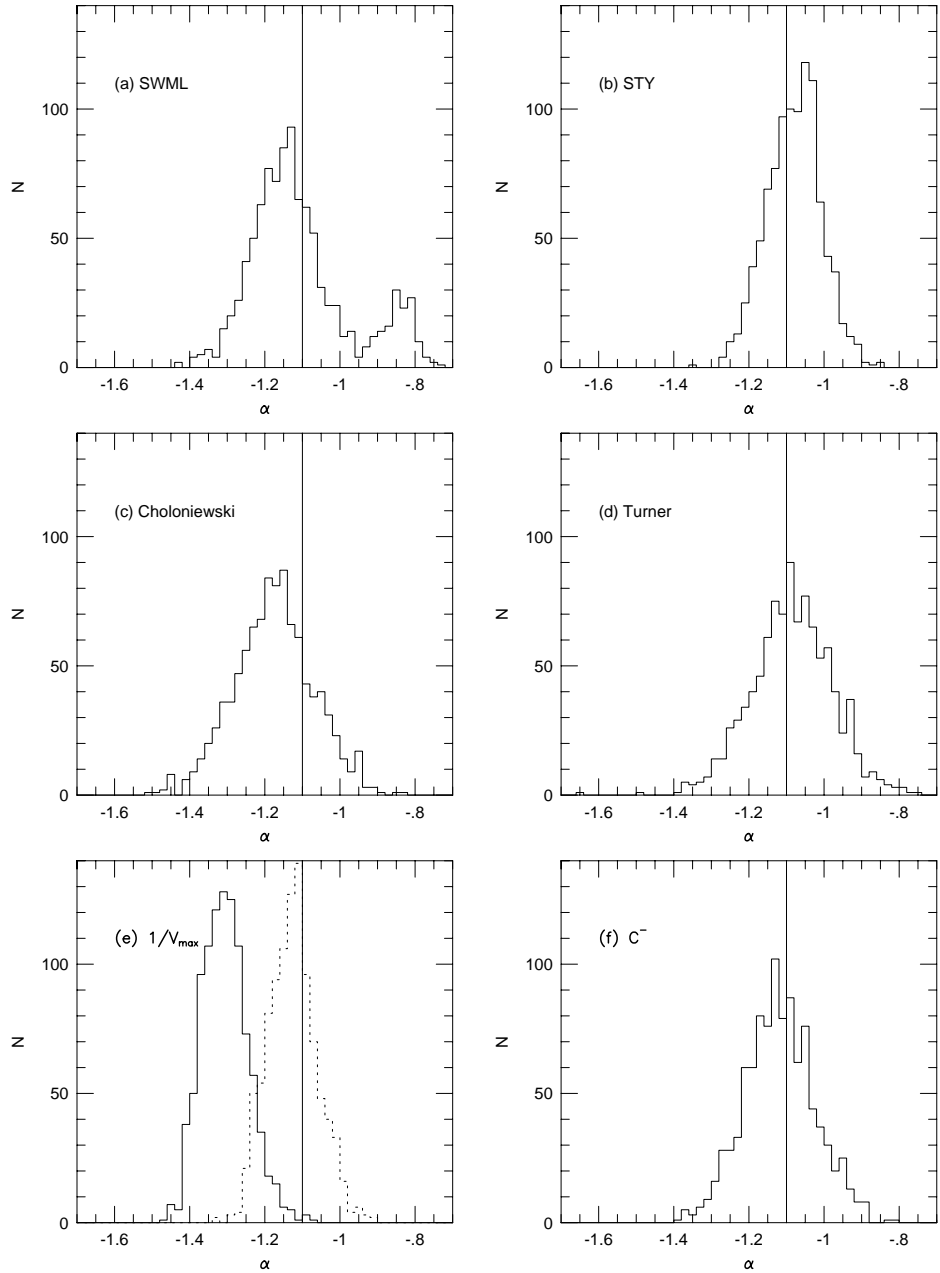}
\figcaption{Histogram showing the distribution of $\alpha$ values, in
0.02 bins. Each panel identifies the method that was used. The
vertical
line indicates the input value of $\alpha$ = -1.10.
For the $1/V_{max}$ estimators, the solid line represents binning in
redshift, while the dashed line represents binning in magnitudes.
}
\end{figure}
\clearpage

\begin{figure}
\vspace{200mm}
\includegraphics{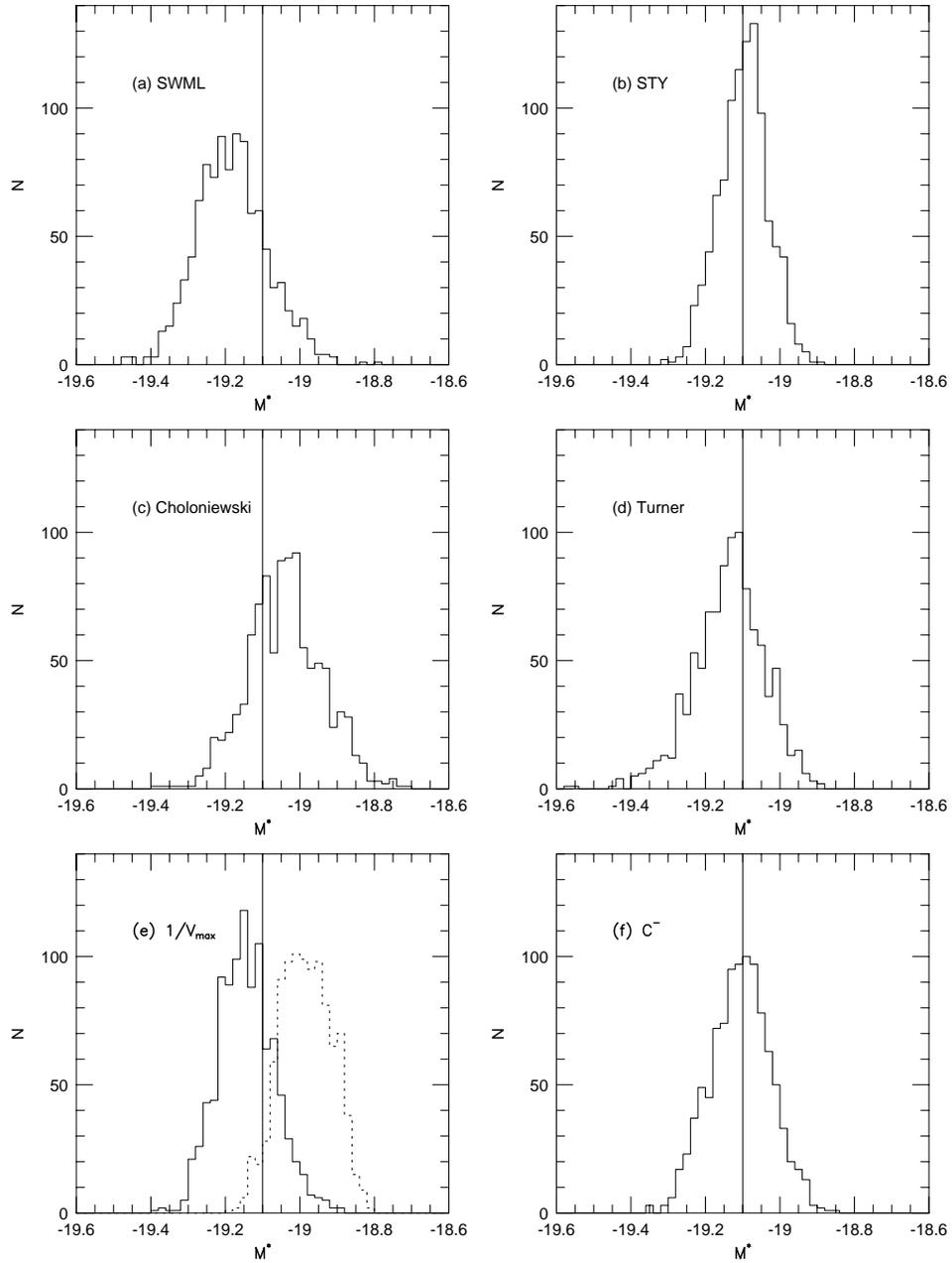}
\figcaption{ Histogram of the $M^*$ values recovered from the
simulations, for each method as shown in the panel. The vertical line
represents the input value of $M^*$. As in the previous figure, for
the $1/V_{max}$ estimators, the solid line represents binning in
redshift, while the dashed line represents binning in magnitudes.
}
\end {figure}
\clearpage

\begin{figure}
\vspace{200mm}
\includegraphics{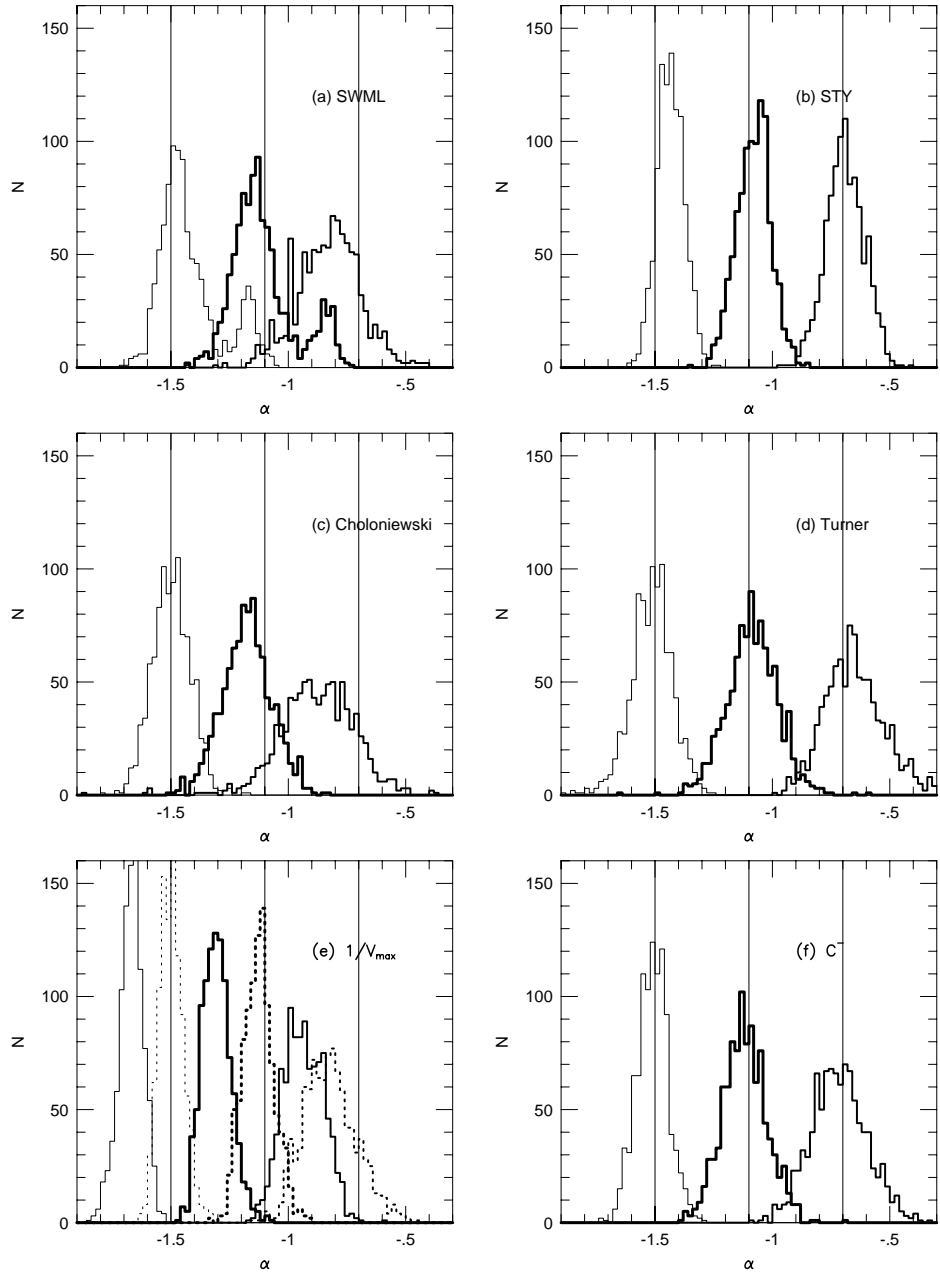}
\figcaption{Histogram comparing the recovered values of $\alpha$ for
three different input values represented as vertical lines (-1.5, -1.1
and -0.7). In the $1/V_{max}$ method, the solid and dotted lines
represent binning galaxies in redshift and magnitudes
respectively. For the sake of clarity, the distribution for each input
value has been coded with a different line weight.
}
\end {figure}
\clearpage

\begin{figure}
\vspace{200mm}
\includegraphics{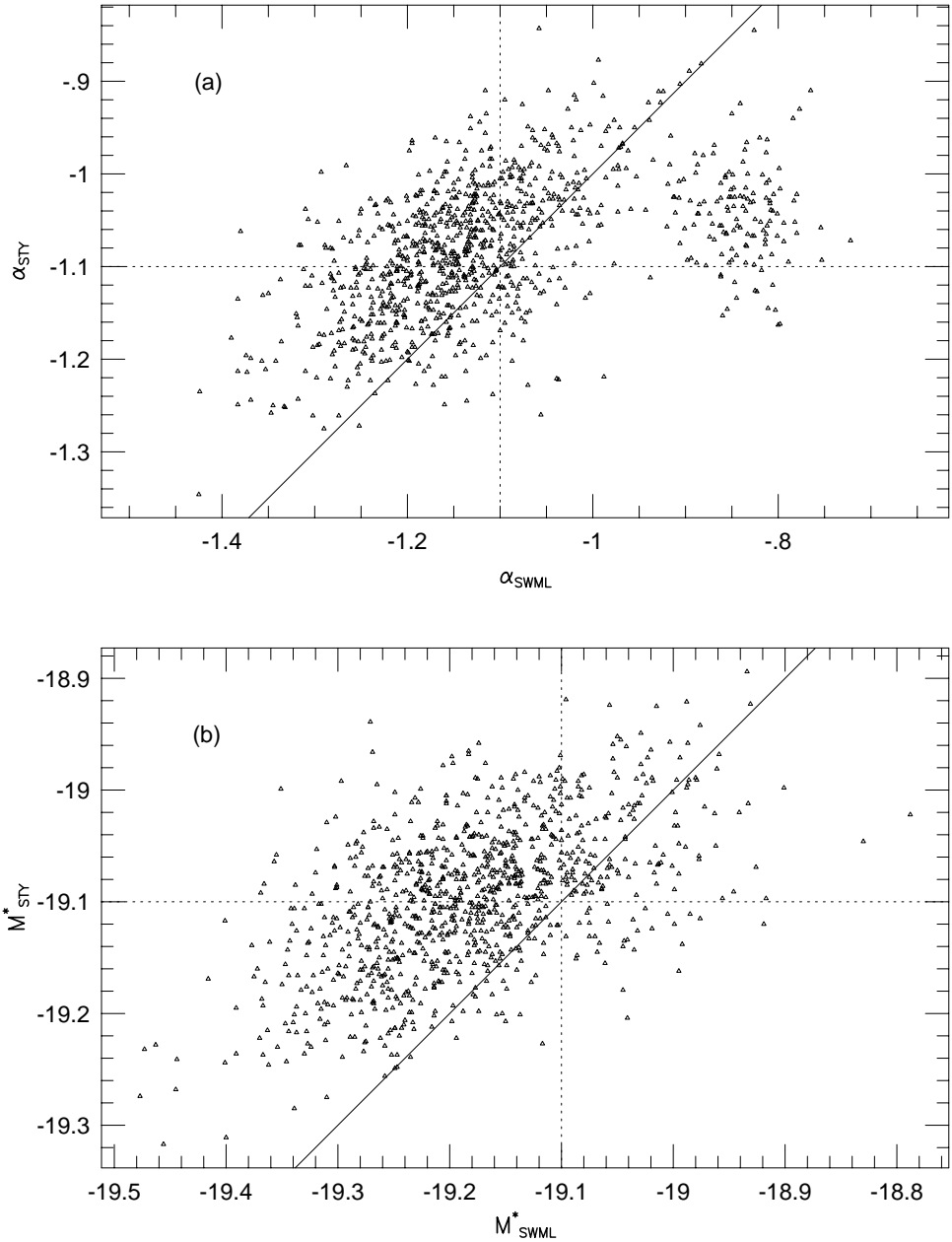}
\figcaption{Comparison between the recovered values for the SWML and STY
methods for $\alpha$ (panel a) and $M^*$ (panel b). The solid line
represents a line with slope equal to 1, while the dotted lines
represent the input values of both parameters.
}
\end {figure}
\clearpage

\begin{figure}
\vspace{200mm}
\includegraphics{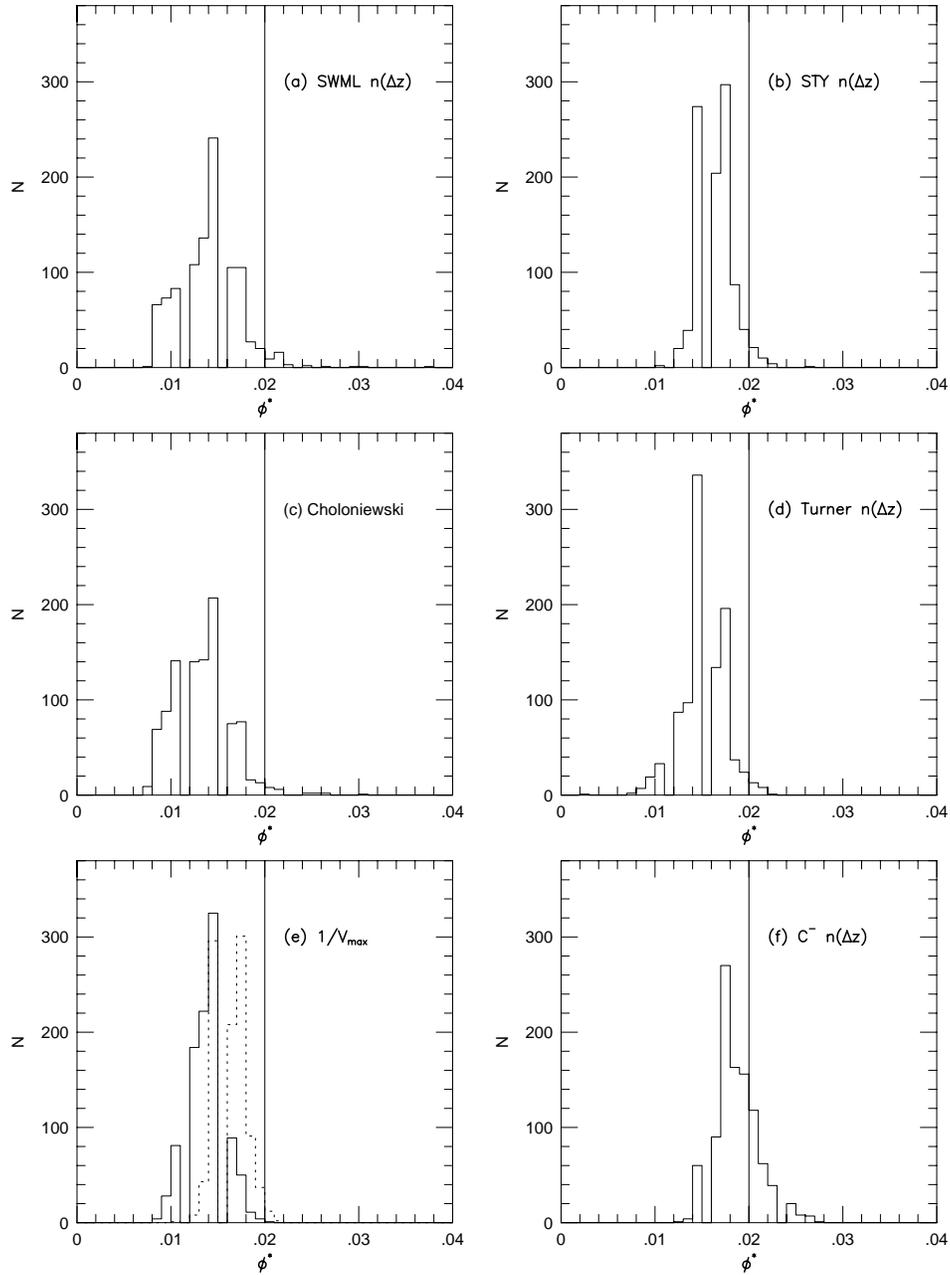}
\figcaption{Histogram comparing a few estimators for the luminosity
function normalization. In this figure we show the estimates obtained
by making three-parameter least-squares fits to the Choloniewski,
and 1/$V_{max}$ methods. For the STY, SWML, Turner and C$^-$ methods we
use the minimum-variance estimator calculated in redshift bins. The
solid vertical line in each panel indicates the input value of $\phi^*$
of the simulations.
}
\end{figure}
\clearpage

%
\begin{figure}
\vspace{200mm}
\includegraphics{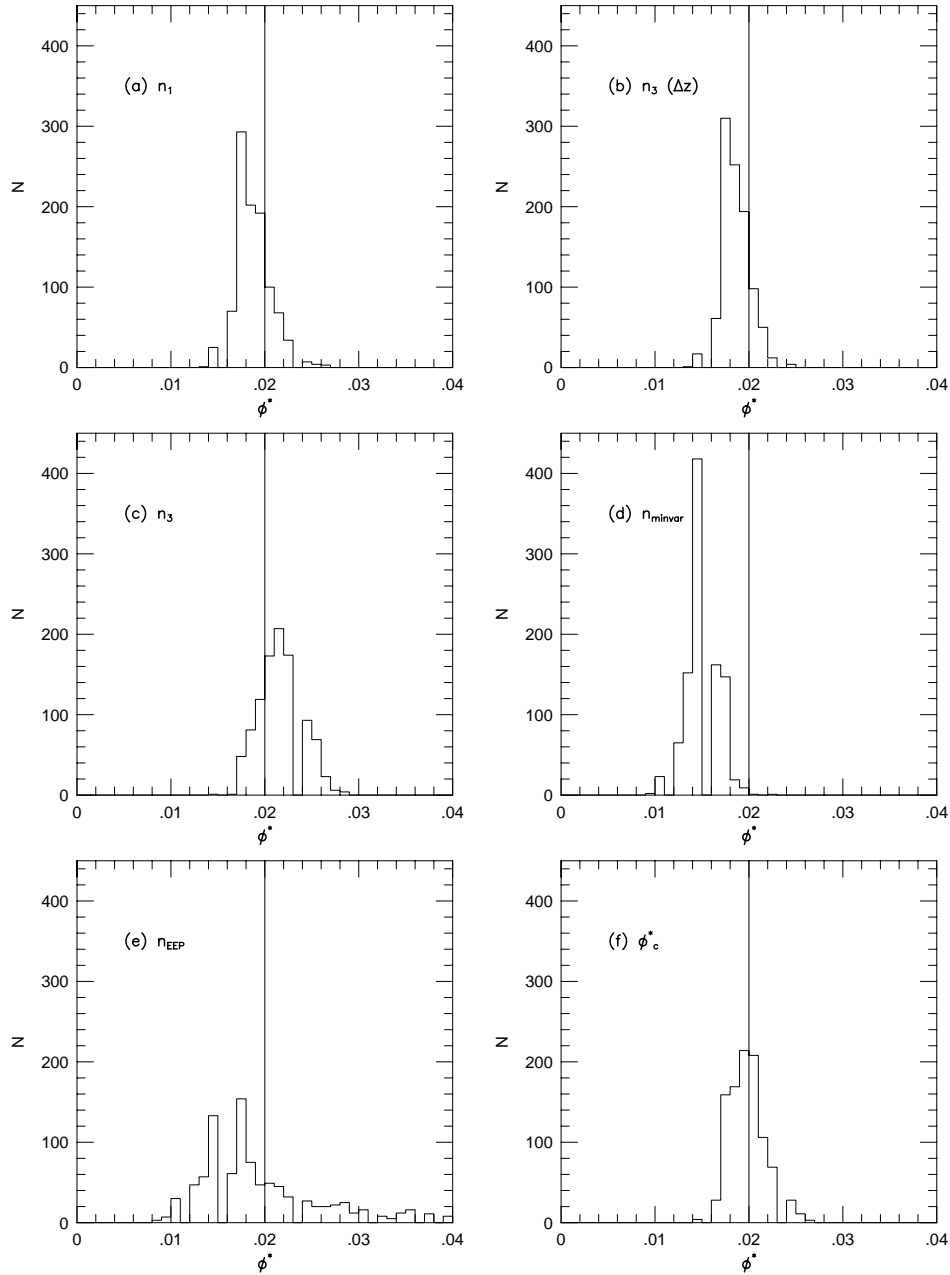}
\figcaption{Histogram comparing six estimators used to calculate $\phi^*$
but in this case only for the STY method. Panel (a) shows the
distribution of $\phi^*$ obtained using the $n_1$ estimator of DH in
redshift bins, while Panel (b) shows the same for the case of
$n_3$. Panel (c) shows another variant of $n_3$, where instead of
counting galaxies in redshift bins, one calculates equation (36) using
the whole surveyed volume. Panel (d) shows the distribution for the
normalization using the minimum variance weighting (equation 34) , but only
considering one galaxy per shell. Panel (e) shows the minimum variance
calculated using equation (38). Panel (f) shows the normalization
using equation (41).
}
\end{figure}
\clearpage

%
\begin{figure}
\vspace{200mm}
\includegraphics{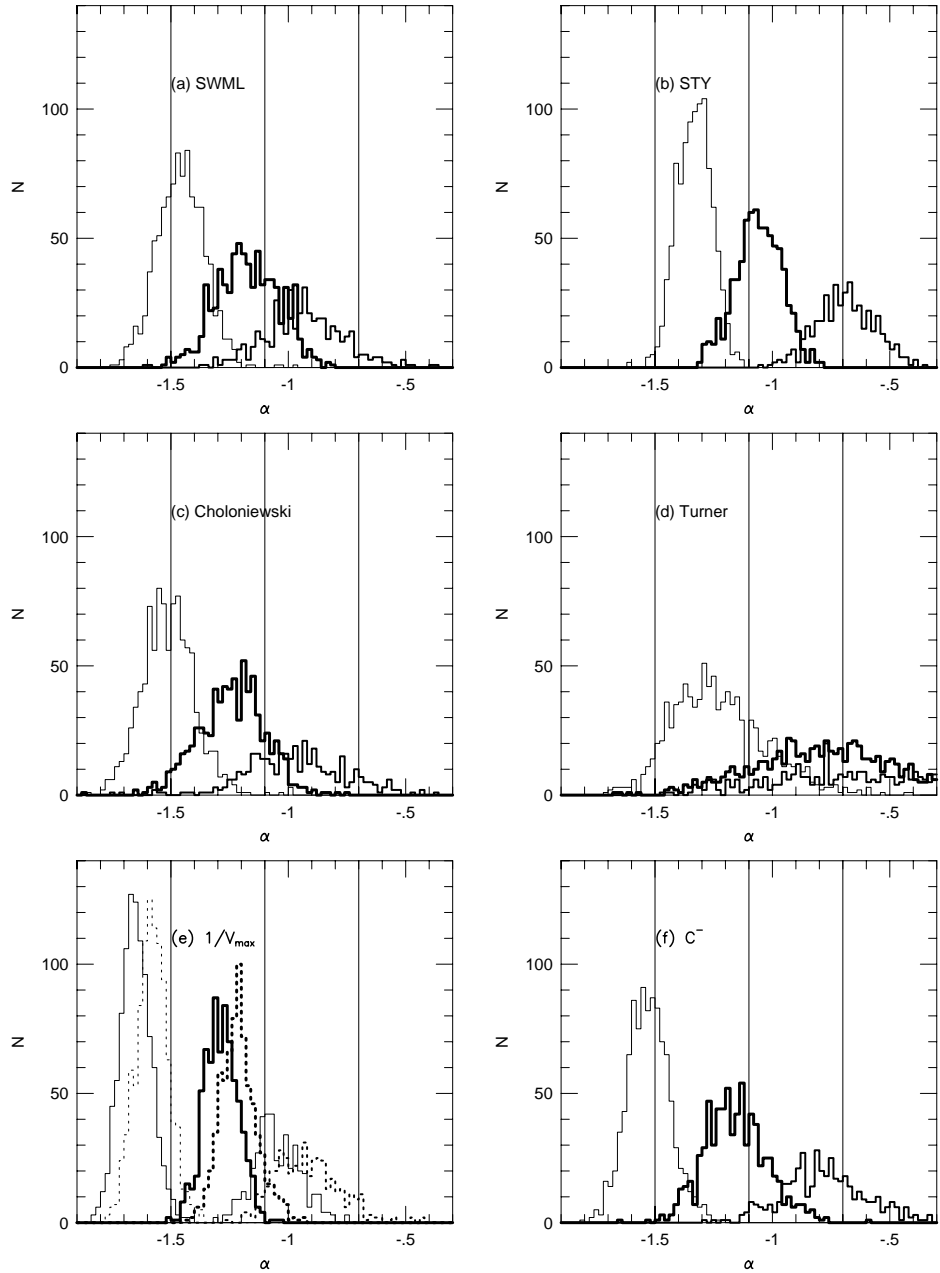}
\figcaption{Histograms showing the recovered $\alpha$ values for samples
with $\sim$ 300 galaxies. As can be seen most methods present a very
wide distribution for lower values of $\alpha$. The full and dotted
lines for the $1/V_{max}$ methods represent binning galaxies in equal
redshift and magnitude bins. As in Fig. 3, the distribution for each
input value has been coded with a different line weight.
}
\end {figure}
\clearpage

\begin{figure}
\vspace{200mm}
\includegraphics{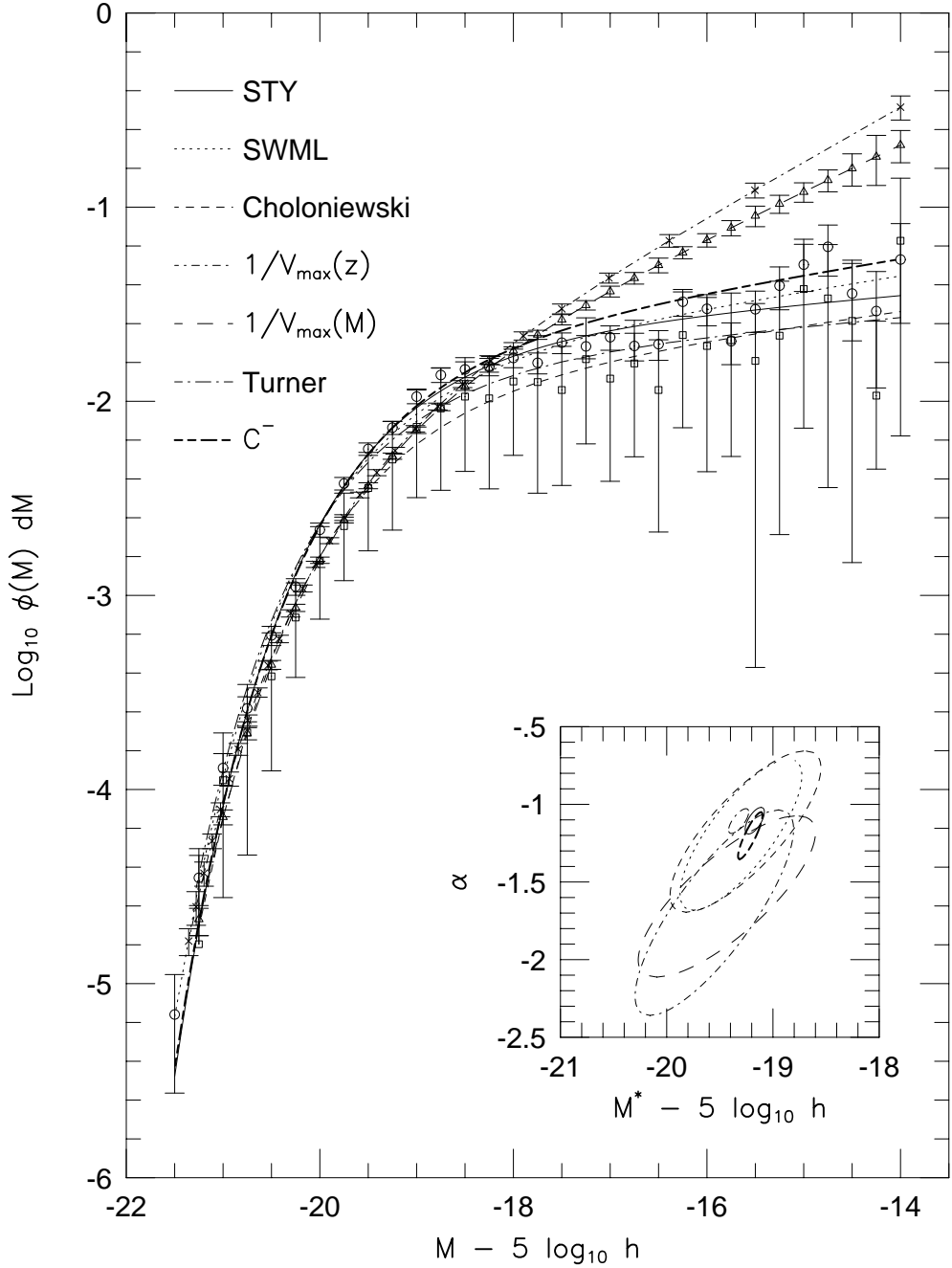}
\figcaption{ A plot showing fits obtained for the CfA1 sample using the
different estimators. Points represent the observed function, while
lines represent the parametric fits. The following estimators are
plotted as symbols: SWML (open circles); Choloniewski  (open squares);
1/$V_{max}$ in redshift bins (crosses) and magnitude bins (open
triangles). The fits for each estimator are coded as noted in the figure. In
the case of the STY, Turner and C$^-$ methods, only the fits are
presented. The inset shows the error ellipses for the fits.
}
\end{figure}
\clearpage

\end{document}